\documentclass[useAMS,usenatbib,dcolumn,usegraphicx,twocolumn,10pt]{mn2e}
\usepackage{amsmath}
\usepackage{placeins}
\usepackage{float}
\usepackage{xspace}

\usepackage[usenames,dvipsnames]{color}
\newcommand{\ud}[1]{\mathrm{d}{#1}}
\newcommand{\br}[1]{\left(#1\right)}
\renewcommand{\sq}[1]{\left[#1\right]}
\newcommand{\kcitet}[1]{\citet{#1}}
\newcommand{\kcite}[1]{\citep{#1}}
\renewcommand{\eqref}[1]{Eq.\ref{#1}}

\renewcommand{\exp}[1]{e^{#1}}

\title[]{Velocity-density twin transforms in thin disk model}
\author[]{
{{\L}ukasz Bratek$^{1}$},
{Szymon Sikora$^{2}$}
{Joanna Ja{\l}ocha$^{1}$},
{Marek Kutschera$^{3}$}
\\
$^{1}$The H. Niewodnicza\'{n}ski Institute of Nuclear Physics,
Polish Academy of Sciences, Radzikowskego 152, PL-31342 Krak\'{o}w, Poland\\
$^{2}$Astronomical
Observatory, Jagellonian University, Orla 171, PL-30244
Krak{\'o}w, Poland\\
$^{3}$The M. Smoluchowski Institute of
Physics, Jagellonian University,  Reymonta 4, PL-30059 Krak{\'o}w, Poland
}

\begin{document}
\date{\today}
\pagerange{\pageref{firstpage}--\pageref{lastpage}} \pubyear{?}

\maketitle

\begin{abstract}
Ring mass density and the corresponding circular velocity in thin disk model are known to be  integral transforms of one another. But it may be less familiar that the transforms can be reduced to one-fold integrals with identical weight functions. It may be of practical value that the integral for the surface density does not involve the velocity derivative, unlike the equivalent and widely known Toomre's formula.
\end{abstract}


\section*{Disk integral transforms}

In this paper we deal with axisymmetric and infinitesimally thin disk model. We use cylindrical coordinate system $\rho,\phi,z$.

Given a surface mass density $\sigma(\rho)$ in the disk plane $z=0$, we can infer the circular velocity $v(\rho)$ of test bodies moving 
in that plane. Conversely, given a  $v(\rho)$, we can 
find the corresponding $\sigma(\rho)$.  
Instead of $\sigma(\rho)$ it is more convenient to consider the ring density $\mu(\rho)=2\pi G\, \rho\, \sigma(\rho)$. 
\noindent{In the next section we show that}
\begin{itemize}
\item[]{ } {\it The quantities $\mu(\rho)$ and $v^2(\rho)$ are related through the following pair of 
mutually inverse integral expressions with the same weighting function $w(x)$:}
\end{itemize}
\begin{eqnarray}\label{main1}
v^2(\rho)&=&\int\limits_{0}^{\infty}
w(x)\,\mu\!\br{x\rho}\,\ud{x},\\
\label{main2}
\mu\br{\rho}&=&\int\limits_{0}^{\infty}w(x)\,
v^2( x^{-1}\rho)\,\ud{x}.
\end{eqnarray} 
 The one-fold integral forms are more
suitable for numerical integration than the equivalent double-integral (or chain forms)  which we shall come to later. 

The weighting function $w(x)$ in both integrals is given by a combination of complete
elliptic integrals $K$ and $E$:\footnote{We use $K$ and $E$ as defined by \kcitet{eliptic}}
$$w(x)=\frac{1}{\pi}\br{ \frac{K\!\sq{k(x)}}{1+x}+
\frac{E\!\sq{k(x)}}{1-x}},\quad k(x)=\frac{2\sqrt{x}}{1+x}.$$
Because $w(x)$ has an integrable singularity at $x=1$, the integration 
should be understood in the Cauchy principal value sense.   The nature of the pole $x=1$ is such that the integrals are sensitive to the radial gradients of the integrands $\mu$ and $v^2$. This is a characteristic feature of the disk model.

\begin{figure}
\centering
\includegraphics[trim =0mm 0mm 0mm 155mm, clip, width=0.9\columnwidth]{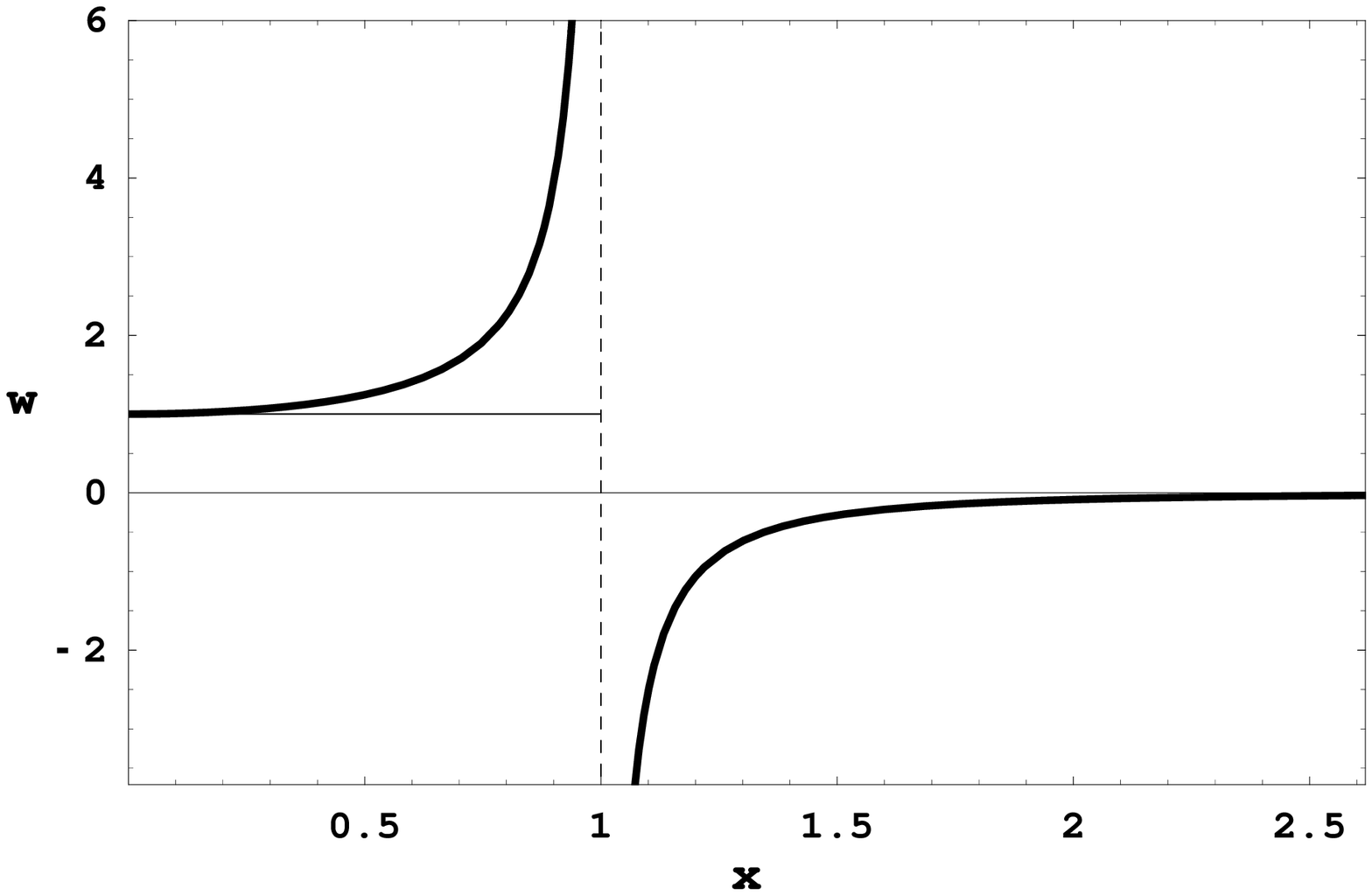}
\caption{Weight function $w(x)$.}
\end{figure}

Despite the wide use of disk model, it seems that \eqref{main2} may not be widely known.
A more familiar is the equivalent \kcitet{Toomre} {\it integrated formula} 
$$
{\sigma}(\rho)\!=\!\frac{G^{-1}}{\pi^2}\! \left[
\int\limits_0^{\rho}\!\frac{\ud{v^2}(\tilde{\rho})}{\ud{\tilde{\rho}}}
\frac{K\!\br{\frac{\tilde{\rho}}{\rho}}}{\rho}\ud{\tilde{\rho}}
+\!\!\int\limits_{\rho}^{\infty}\!\frac{\ud{v^2}(\tilde{\rho})}{\ud{\tilde{\rho}}}
 K\!\br{\frac{\rho}{\tilde{\rho}}}\!\frac{\ud{\tilde{\rho}}}{\tilde{\rho}}\right]\!.
$$which contains only a logarithmic singularity of $K$,  but on the cost of 
involving the derivative of $v^2$ which introduces additional uncertainty in modeling galactic disks.
The usual advice about numerical evaluation of 
singular integrals is to integrate by parts, which may be the reason why \kcitet{Toomre} gave no expression in form of a one-fold integral without the derivative of
$v^2$. But the singularity in $w(x)$ is nowadays easily tractable numerically and presents itself no difficulty at all. 

A word of
warning may be appropriate, here.
Owing to the sensitivity to radial gradients mentioned above, the disk model was pointed out in the context of  Toomre's integrated formula to be of relatively little use in practice on account of the fact that the derivative of $v^2$ is usually subject to significant observational errors, resulting in a $\sigma$ varying in an erratic and unphysical way \kcite{Binney}.
The disk model has also other limitations. In realistic situations the rotation curves do not extend far enough 
and one has to extrapolate the data. Unfortunately, the results 
depend on the way one choses to extrapolate. Therefore, the disk model must be used with due care.

We decided to focus on the integral form \eqref{main2} in this separate paper, because of usefulness of a formula turning the rotation curve to the surface density in modeling galactic disks. A reduced one-fold integral form is needed for practical reasons, for the accuracy and speeding up the numerical integration, especially in finding column mass densities of finite-width disks by means of recursions, like in \kcite{Jalocha14}.  

There are also known various forms of equivalent double integral representations of \eqref{main2}, e.g. \kcite{method1} or those implied by \kcitet{Toomre} or  \kcitet{Kalnajs} methods  which we focus later on. We  recall also that
there are numerical methods of finding $\sigma$ from a fragment of $v^2$, given some other measurements complementary to the rotation data \kcite{Jalocha}. 
An algebraic approach to inverting \eqref{main1} presented by \kcitet{method2}, offers an interesting alternative to the direct formula \eqref{main2} represented on a union of osculating rings, if it can be assumed that $\sigma$ practically  vanishes beyond the last measured point of $v$. With transforms \eqref{main1} and \eqref{main2} cut-off at the last point, the same result can be then obtained by iterations, analogous to those in  \kcite{Jalocha}, assuming vanishing density beyond the cutoff.

\section{\label{sec:deriv}Twin transforms from Toomre's method}

Surface density for axi-symmetric discs is naturally expressed in terms of Hankel transforms 
which are a 
special case of Fourier transforms involving circular symmetry. For our purposes we intentionally rewrite the result of \kcitet{Toomre} method into the chain form corresponding to \eqref{main2}
$$\mu(\rho)=\int\limits_0^{\infty}\br{\int\limits_0^{\infty}{\lambda}
\,\rho J_0\br{\lambda\rho}\tilde{\rho}J_1(\lambda\tilde{\rho}) \,\ud{\lambda}  }
{v^2(\tilde{\rho})}\,\frac{\ud{\tilde{\rho}}}{\tilde{\rho}}.$$
Toomre called it as {\it  too formal to be of any direct use} and, having integrated by
parts, gave his integrated formula  as
the final result. 
Nevertheless, the above form with Bessel functions  is useful in finding analytical expressions for $\sigma$, given a $v^2$ (or vice versa), e.g. \kcite{Freeman}.

\medskip\noindent
In order to prove \eqref{main2}, we need to calculate the integral in the round brackets of the above chain form.  
The inverse chain form corresponding to \eqref{main1} can be easily deduced, e.g. \kcite{Bratek08}, and we arrange the result into a form resembling the previous integral
$$
{v^2(\rho)}=\int\limits_0^{\infty}
\underbrace{ 
\br{\int\limits_0^{\infty} {\lambda} \,\tilde{\rho} J_0(\lambda\tilde{\rho})\,\rho J_1(\lambda\rho)\,\ud{\lambda}
}
}_{\equiv {T}(\rho,\tilde{\rho})}\mu(\tilde{\rho})\,\frac{\ud{\tilde{\rho}}}{\tilde{\rho}}.
$$
It is evident the symmetry
${v^2(\rho)}=\int_0^{\infty}T(\rho,\tilde{\rho})\mu(\tilde{\rho})\,\frac{\ud{\tilde{\rho}}}{\tilde{\rho}}$,
$\mu(\rho)=\int_0^{\infty}T(\tilde{\rho},\rho)
{v^2(\tilde{\rho})}\,\frac{\ud{\tilde{\rho}}}{\tilde{\rho}}$. 
By substituting $\tilde{\rho}=x\rho$ in the first integral,
we obtain
${v^2(\rho)}=\int_0^{\infty} T(\rho,x\rho)\mu(x\rho)\,\frac{\ud{x}}{x}$, whereas   substituting $\tilde{\rho}=\rho/x$ in the second integral we obtain
$\mu(\rho)=\int_0^{\infty}T(\rho/x,\rho) 
{v^2(\rho/x)}\,\frac{\ud{x}}{x}$. 
Furthermore, it is easily seen that
$T(\rho_1,\rho_2)=\frac{\rho_2}{\rho_1}u\br{\frac{\rho_2}{\rho_1}}$, where
${ u(x)\equiv\int_0^{\infty}\omega J_0(\omega x)J_1(\omega)\,\ud{\omega}}$.
As so, ${v^2(\rho)}\!=\!\!\int_0^{\infty}\!\!u(x)\mu(x\rho)\,\ud{x}$ and $\mu(\rho)\!=\!\!\int_0^{\infty}\!\!u(x)
v^2(\rho/x)\,\ud{x}$, which explains why the weight function in \eqref{main1} and \eqref{main2} are identical.

 To complete our derivation, it remains to determine
$u(x)$. Instead of using tables of integrals, we can  deduce $u(x)$
by comparing the previous expression for $v^2$ with one from a textbook calculation concerning the axisymmetric gravitational potential $\Phi(\rho,z)$ of a thin disk. 
First, we arrange the expression for $\Phi$ so as to isolate the elliptic function $K$  (for a fixed $\phi$ we make use of  a new integration variable  $\tilde{\phi}\to\gamma$: $2\gamma=\tilde{\phi}-\phi+\pi$) $${ \Phi\!\br{\rho,z}=\!-4G\!\!\int\limits_{0}^{\infty}\!\!
\frac{\tilde{\rho}\, \sigma(\tilde{\rho})\,\ud{\tilde{\rho}}}{\sqrt{(\rho+\tilde{\rho})^2+z^2}}\!\int\limits_{0}^{\pi/2}\!\!\!\!\frac{\,\ud{\gamma}}{\sqrt{1-\!\frac{4\rho\tilde{\rho}}{
(\rho+\tilde{\rho})^2+z^2}\sin^2{\gamma}}}.}$$  
By differentiating $\Phi$ with respect to $\rho$,
using the property $K'(k)=\frac{E(k)}{k(1-k^2)}-\frac{K(k)}{k}$ and
 taking the limit $z\to0$, we can obtain the desired result from the force equilibrium condition $\rho^{-1}v^2(\rho)=\partial_{\rho}\Phi\br{\rho,0^{\pm}}$ for circular orbits in the disk plane. By substituting $\tilde{\rho}=\rho x$, and denoting $k(x)\equiv{2\sqrt{x}}/\br{1+x}$, the result can be simplified to
$$\label{eq:inv}\frac{v^2\br{\rho}}{\rho}=2G\int\limits_{0}^{\infty}
\br{
\frac{K\!\br{k(x)}}{1+x}+\frac{E\!\br{k(x)}}{1-x}}\sigma\!\br{\rho\,x}x\,\,\ud{x}.
$$
From this result we immediately see that $u(x)\equiv w(x)$, which in turn proves the relation \eqref{main2}.\footnote{A similar expression to \eqref{main2} we obtained in a not so straightforward way in \kcite{bratek} and it is connected with the present form by the inversion $x\to x^{-1}$.}

\section*{Twin transforms from Kalnajs' method}
There is a more sophisticated way of understanding the fact that
 the integrals turning $v^2$ to $\mu$ and  $\mu$ to $v^2$ can be put into forms with the same weight function. 

\kcitet{Kalnajs} related column 
density $\sigma(\rho)$ to the circular velocity $v^2(\rho)$ in the plane $z=0$ for a spheroid with similar isodensity surfaces
$\rho^2+z^2/q^2=m^2$ characterized by a fixed flattening $q$ and parameterized by $m$. 
With the definitions $\hat{P}(\alpha)\equiv\frac{\Gamma(3/2)\Gamma((1-i\alpha)/2)}{\Gamma(1-i\alpha/2)}$ and $\hat{S}(\alpha,q)\equiv\br{1+i\alpha}^{-1}\cdot {}_2F_1(1,1+i\alpha,(3+i\alpha)/2,(1-q)/2)$,  Kalnajs' result can be arranged in the chain form
$${v^2}(\rho_o\exp{u})=\!\!\int\limits_{-\infty}^{\infty}\!\!\br{\frac{1}{2\pi}\!\!\int\limits_{-\infty}^{\infty}\!\!\frac{\hat{S}(\alpha,q)}{\hat{P}(\alpha)}\,\exp{i\alpha(u-\tilde{u})}\ud{\alpha}  }
{{\mu}(\rho_o\exp{\tilde{u}})}\,\ud{\tilde{u}},$$
Kalnajs' method exploits the translational symmetry in the scale-invariant variable $u=\ln\br{\rho/\rho_o}$ ($\rho_o$ being an arbitrary and fixed scale parameter).
For $q=0$ the ratio  $\frac{\hat{S}(\alpha,q)}{\hat{P}(\alpha)}$ reduces to $\frac{\hat{P}(-\alpha)}{\hat{P}(\alpha)}$ with absolute value $1$. It  then makes sense to consider the inverse convolution whose Fourier transform is $\frac{\hat{P}(\alpha)}{\hat{P}(-\alpha)}$.\footnote{
For nonzero flattening, the ratio $\frac{\hat{S}(\alpha,q)}{\hat{P}(\alpha)}$ tends to $0$ at infinity and we cannot write  the inverse convolution form, the absolute value of that ratio only
 tends non-uniformly to $1$ as $q\to0$. } This allows us to write down  the following integrals with identical weight functions 
\begin{eqnarray*}v^2(\rho_o\exp{u})&=&\!\!\int\limits_{-\infty}^{\infty}\br{\frac{1}{2\pi}\!\!\int\limits_{-\infty}^{\infty}\!\! \frac{\hat{P}(-\alpha)}{\hat{P}(\alpha)}\,\exp{i\alpha (u-\tilde{u})}\ud{\alpha} }
{\mu(\rho_o\exp{\tilde{u}})}\,\ud{\tilde{u}}
,\qquad\\ \mu(\rho_{o}\exp{-u})&=&\!\!\int\limits_{-\infty}^{\infty}\br{\frac{1}{2\pi}\!\!\int\limits_{-\infty}^{\infty}\!\! \frac{\hat{P}(-\alpha)}{\hat{P}(\alpha)}\,\exp{i\alpha (u-\tilde{u})}\ud{\alpha} }
{v^2(\rho_o\exp{-\tilde{u}})}\,\ud{\tilde{u}},\end{eqnarray*} 
 (in the second integral we have reflected the variables $\alpha, u, \tilde{u}$ with respect to 
$0$). The above expressions are  counterparts  of Toomre's chain forms.

\medskip
As a byproduct from the two methods we can deduce  the following integral representations of function $w(x)$:
$$w(x)=\left\{
\begin{array}{l}
{ \frac{1}{2\pi}\!\!\int\limits_{-\infty}^{\infty} \frac{\Gamma\br{\frac{1}{2}+\frac{i\alpha}{2}}\Gamma\br{1-\frac{i\alpha}{2}}}{\Gamma\br{\frac{1}{2}-\frac{i\alpha}{2}}\Gamma\br{1+\frac{i\alpha}{2}}}\,\exp{-(1+i\alpha)\ln{x}}\,\ud{\alpha}},
\\{}\\
{\int\limits_0^{\infty}\omega J_0(\omega x)J_1(\omega)\,\ud{\omega}}.
\end{array}\right.
$$

\newcommand{\mnras}{MNRAS}
\newcommand{\apj}{ApJ}
\newcommand{\aap}{A\&A}
\bibliography{transforms}
\bibliographystyle{mn2e} 

\end{document}